# Production of $^{177}$Lu with deuterons at IFMIF-DONES facility

Elena López-Melero[1], Francisco García-Infantes[1,2], Isabel López-Casas[1], Fernando Arias de Saavedra[1], Ignacio Porras[1], Andrés Roldán[3], Laura Fernández Maza[4], Javier Praena[1,2*]

[1]Departamento de Física Atómica, Molecular y Nuclear, Universidad de Granada, Spain
[2]European Organization for Nuclear Research (CERN), Geneva, Switzerland
[3]Departamento de Electrónica y Tecnología de Computadoras, Universidad de Granada, Spain
[4]Hospital Virgen de la Arrixaca, Murcia, Spain

### Abstract

The International Fusion Materials Irradiation Facility - Demo Oriented NEutron Source (IFMIF-DONES) is a single-sited novel Research Infrastructure for testing, validation and qualification of the materials to be used in a fusion reactor. The main purpose of IFMIF-DONES is related to fusion technology and neutron irradiation of the materials to be used in the future fusion power plants. However, there is an important effort to take advantage of the outstanding characteristics of the facility in terms of neutrons and deuterons. One of the applications could be radioisotopes production with deuterons and neutrons. We discuss here the possible production of radioisotopes with deuterons at DONES. In this work, we have focused on the production of $^{177}$Lu with deuterons. The study has been carried out through the design and simulation of a device as cooling system for the sample producing the radioisotope. The results show the viability of using DONES for such production. In addition, the study suggests that new nuclear energy data above 20 MeV for deuterons is mandatory for an accurate study in this field.

## 1. INTRODUCTION

IFMIF will be an installation for fusion material irradiation that will simulate the conditions inside a fusion reactor in terms of neutron flux and neutron energy [1]. The nuclear reaction between deuterium and lithium nucleus was selected as the most suitable for simulating the neutron emission from the deuterium-tritium reaction in the future fusion reactor DEMO. Thus, together with ITER (International Thermonuclear Experimental Reactor), IFMIF is a key facility to design DEMO reactor. The construction of an 'Early DEMO' was decided, DONES (DEMO-Oriented NEutron Source), which basically consists of a simplification of IFMIF. The IFMIF project will consist of two accelerators, meanwhile, DONES will consist of a single accelerator.

Granada (Spain) was selected as European city host for the construction of the facility and IFMIF-DONES was selected as a key energy infrastructure, by ESFRI (European Strategy Forum on Research Infrastructures) [2]. At present, the IFMIF-DONES Preparatory Phase project deals with the financial, legal and organisational issues related to the international character of the facility during its construction and operation phases [3].

The key infrastructures of the facility are the accelerator and the lithium target for neutron production [4]. The planned accelerator to be included in IFMIF-DONES will accelerate with a current of 125 mA deuterons up to 40 MeV, which means that it must handle a power of 5 MW.

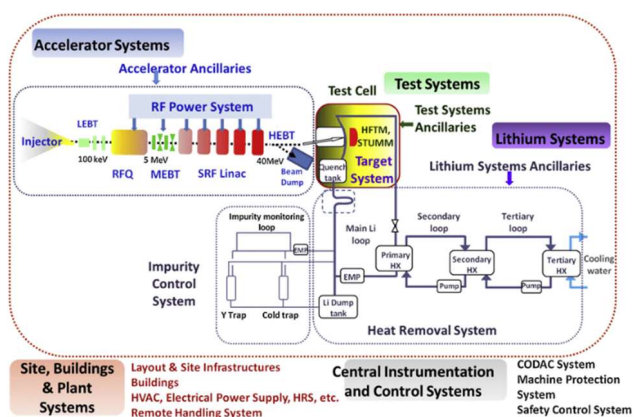

Figure 1. General scheme of operation of IFMIF-DONES. The accelerator, the target and the test cell [4].

The deuteron beam will strike the liquid lithium target that needs to circulate at high speed, 15 m/s, to extract the high incident power. IFMIF-DONES will reach a neutron flux of $10^{18}$ m$^{-2}$s$^{-1}$ with a broad peak at 20 MeV. Fig. 1 shows a simply scheme of IFMIF-DONES [5]. The key materials most exposed in DEMO will be irradiated in the high flux test module (HFTM) and in the other two test cells less exposed.

Although IFMIF-DONES has a clear and an important objective, the characteristic of the facility allows to study possible applications beyond the study of materials. The White Paper included a preliminary report on a complementary scientific program at IFMIF-DONES [6] with several applications from fundamental physics to chemistry and biology.

In a further report, we studied the use of IFMIF-DONES for testing electronic devices under neutron and photon irradiation, preliminary calculations on neutron scattering and preliminary calculations on the production of radioisotopes with deuterons for nuclear medicine [7].

The deuteron beam at 40 MeV and 125 mA provides unique possibilities in radioisotope production. The present work will be focused on the main constrain related with the production: the very high power to be sustained by the sample producing the radioisotope. Thus, a device will be studied with the objective to sustain the highest power. A simulation of its behaviour will determine the maximum activity and specific activity that it could be produced in realistic conditions. As a first case, the production of $^{177}$Lu is considered.

According to the tabulated data of the IAEA [8], the average emission energies of electrons for $^{177}$Lu are in the range [40-150] keV. This range makes the radioisotope suitable for therapeutic use. On the other hand, during the disintegration the gamma radiation causes it to be detected with imaging techniques. For it, $^{177}$Lu is used for theranostics (therapy and diagnosis). As a β emitter-radiopharmaceutical, it can be used to treat neuroendocrine tumours ($^{177}$Lu-DOTATATE) [9]; and more recently, $^{177}$Lu has been included in clinical trials for the treatment of castration-resistant metastatic prostate cancer ($^{177}$Lu-PSMA) [10]. $^{177}$Lu-Anti HER2 seems to be a potential theranostic agent for breast cancer [11]. $^{177}$Lu chemical properties allow its binding to antibodies, which opens a wide range of new theranostic radiopharmaceuticals [12].

Currently, $^{177}$Lu ($T_{1/2}$=6.65 d) is only produced in nuclear reactors with neutron capture on Lu or Yb enriched samples: $^{176}$Lu(n,γ)$^{177(m+g)}$Lu and $^{176}$Yb(n,γ)$^{177}$Yb [13]. It should be noticed that the production of $^{177m}$Lu is undesirable due to its long life that delivers unwanted dose to the patient. Also, it complicates the post clinical follow up of the patient because urine must be treated as radioactive waste.

The production with deuteron beam is possible with the following two routes:

- Direct route: d+$^{176}$Yb→n+$^{177(m+g)}$Lu
- Indirect route: d+$^{176}$Yb→p+$^{177}$Yb→$^{177g}$Lu

In the direct route, the produced nuclei are the same that in the neutron capture on $^{176}$Lu. However, the deuteron production of the metastable state would be negligible [14]. In the indirect route, the produced nucleus is the same that in the neutron capture on $^{176}$Yb, therefore, free of the undesirable $^{177m}$Lu.

In previous a work, we carried theoretical studies on the production of $^{177}$Lu, to have a first approximation to the results that can be obtained in IFMIF-DONES on the production of $^{177}$Lu [14]. In the present work, we design a configuration closer to reality and obtain the production of radioisotopes with deuterons in IFMIF-DONES, a model of a millichannels cooling system has been proposed, the objective consists of a sample of $^{176}$Yb$_2$O$_3$ deposited on a copper backing cooled by running water. In addition, simulations have been performed using SolidWorks to obtain a final model. In the next section we will study the production of $^{177}$Lu with a sample of $^{176}$Yb through the two mentioned routes that would be obtained in IFMIF-DONES.

## 2. MATERIALS AND METHODS

The accelerator of IFMIF-DONES opens two production routes to produce $^{177}$Lu, which are interesting to study. Production rate, activity, and mass of lutetium with both production routes will be studied.

The production of lutetium, through the direct route, is obtained from the differential equation:

$$\frac{dN_{^{177}Lu}}{dt} = R_{^{177}Lu} - N_{^{177}Lu}\lambda_{^{177}Lu} \qquad (1)$$

where $dN_{^{177}Lu}$ is the number of nuclei produced, $R_{^{177}Lu}$ is the rate of production and $\lambda_{^{177}Lu}$ is the half-life. The production rate is given by:

$$R_{^{177}Lu} = \frac{I}{q}\int_{E_f}^{E_i}\frac{\sigma(E)}{S_{Yb}^d(E)}dE \qquad (2)$$

where $\sigma(E)$ is the cross-section of the reactions, $S_{Yb}^d(E)$ is the stopping power of the deuterons, I is the intensity of the incident beam and q is the charge of the incident particle.

Therefore, the number of nucleus produced by the direct route remains:

$$N_{^{177}Lu} = \frac{I(1-e^{-\lambda_{^{177}Lu}t})}{q\lambda_{^{177}Lu}}\int_{E_f}^{E_i}\frac{\sigma_{dir}(E)}{S_{Yb}^d(E)}dE \qquad (3)$$

For the indirect route, the differential equation must consider the time of disintegration of the ytterbium. In this way, the differential equation to solve is:

$$\frac{dN_{^{177}Lu}}{dt} = R_{^{177}Lu}\left(1 - e^{-\lambda_{^{177}Lu}t}\right) - N_{^{177}Lu}\lambda_{^{177}Lu} \qquad (4)$$

Note that the cross-sections of both reactions are different. The number of nucleus produced remains:

$$N_{^{177}Lu}(t) = \frac{I}{q}\int_{E_f}^{E_i}\frac{\sigma_{ind}(E)}{S_{Yb}^d(E)}dE\left(\frac{\left(1-e^{-\lambda_{^{177}Lu}t}\right)}{\lambda_{^{177}Lu}}+\right.$$
$$\left.\frac{e^{-\lambda_{^{177}Yb}t}-e^{-\lambda_{^{177}Lu}t}}{\lambda_{^{177}Yb}-\lambda_{^{177}Lu}}+\frac{1-\exp^{-\lambda_{^{177}Yb}t}}{\lambda_{^{177}Lu}-\lambda_{^{177}Yb}}\right) \qquad (5)$$

Now we must solve the integrals of expression (3) and expression (5) [8]. For it, we are going to analyse each parameter of the equations in detail.

## 2.1 STOPPING POWER

The stopping power is given by the semi-empirical expression developed by Ziegler and Andersen. This expression was developed for protons, and is given by [15]:

$$S_{Yb}^{p}(E) = \frac{A_6}{\beta^2}\left[\ln\left(\frac{A_7\beta^2}{1-\beta^2}\right) - \beta^2 - \sum_{i=0}^{4} A_{i+8}\,(\ln(E))^i\right] \tag{6}$$

where $\beta = \frac{v}{c}$, E is the energy of the incident beam and the values of $A_i$ is a particular constant of the ytterbium.

Considering that the stopping power depends on the energy and speed of the incident particle, the relation between the expression for the stopping power of protons and deuterons is:

$$S_{Yb}^{p}(E) = S_{Yb}^{d}\left(\frac{m_p}{m_d}(E)\right) \tag{7}$$

where $m_p$ and $m_d$ are the mass of a proton and deuteron, respectively. The same expressions are used by the SRIM code (Stopping and Range of Ions Matter) [16]. We will use SRIM for a double check of the energy losses by the deuterons in the target of Yb for different thicknesses.

## 2.2 CROSS SECTION

For the values of the cross-section, the experimental data provided by Hermanne et al. [17] and Manenti et al. [18] have been taken. These data range up to 20 MeV, so for higher energies we must rely on theoretical fits of the cross-sections, see Fig. 2. For the indirect route, the fit is (red line):

$$\sigma_{ind}(E) = \begin{cases} \frac{d}{s\sqrt{2\pi}}e^{\frac{-(E-u)^2}{2s^2}}, & 0 < E < 11.22\ MeV \\ n \cdot e^{-\frac{E}{k}}, & 11.22\ MeV \le E < 40\ MeV \end{cases} \tag{8}$$

where d=(1.66±0.07)·10^6 mb·keV, s=(2.72±0.16)·10^3 keV, u=(1.16±0.11)·10^4 keV, k=(1.19±0.08)·10^4 keV and n=(620±50) mb [17].

On the other hand, the setting for the fit of the direct route is (dark line):

$$\sigma_{dir}(E) = a + b \cdot E \tag{9}$$

where a=(-54±5) mb and b=(4.4±0.3)·10^-3 mb/keV [18]. Fig. 2 shows the fits performed up to 40 MeV for both cross-sections based on the method of our work.

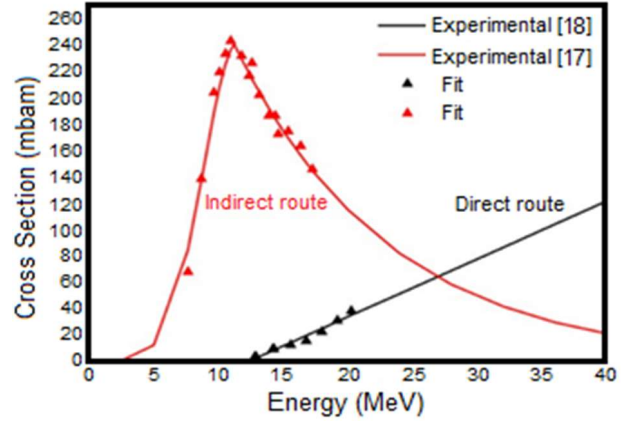

Figure 2. Cross sections (mbarn) of the experimental data (dots) and theoretical fit (lines) obtained by following Arias de Saavedra et al. [19].

## 2.3 ACTIVITY AND SPECIFIC ACTIVITY

Finally, once we have optimized all the parameters, we calculate the activity, mass and specific activity of $^{177}$Lu.

The activity of the lutetium sample is given by the expression:

$$A(t) = \lambda_{^{177}Lu}N_{^{177}Lu}(t) \tag{10}$$

where $N_{^{177}Lu}(t)$ is the number of nuclei generated of $^{177}$Lu, which is obtained by adding the result of the expressions 3 and 5. On the other hand, the produced mass of lutetium is given by the expression:

$$m_{^{177}Lu}(t) = N_{^{177}Lu}(t) \cdot \frac{176.94\ g}{6.022 \cdot 10^{23}} \tag{11}$$

Finally, having the activity and mass of lutetium after irradiation, we can calculate the specific activity.

## 3. BACKING DESIGN WITH SOLIDWORKS

In order to perform a realistic study on power dissipation and maximum activity produced, several simulations have been carried out with SolidWorks.

The variables taken into account for these simulations have been: the inlet fluid velocity, the distance between the $^{176}$Yb$_2$O$_3$ sample and the millichannels and the separation between millichannels, which improves or not heat dissipation.

Different velocities were tested for the fluid, the goal was to find an engagement such that the velocity was high enough to dissipate heat and as low as the backing wouldn't burst. Focusing on this, the velocity which fits best with these requirements is 12 m/s.

Once the fluid velocity was set at 12 m/s, the other parameters have been varied, distance between the $^{176}Yb_2O_3$ sample and the millichannels and separation between millichannels, until finding the setup which allows having the maximum thick of Cu without reaching its melting temperature. Fig. 3 shows a temperature map as function of our parameters. Remember, Cu melting temperature is 1358 K.

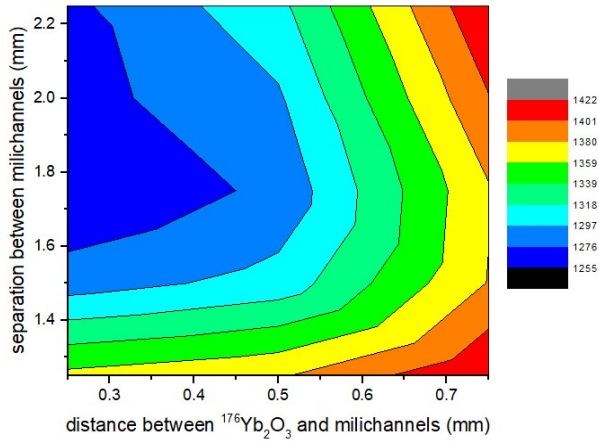

Figure 3. Map of temperature as function of the distance between the $^{176}Yb_2O_3$ sample and the millichannels and the separation between millichannels. Cu melting temperature is 1358 K, this means that from the included yellow area, our backing melts.

Thus, as the yellow area in Fig. 3 establishes Cu melting point, the values chosen for the distance between the $^{176}Yb_2O_3$ sample and the millichannels and the separation between millichannels have been 0.5 mm and 1.5 mm, respectively. These parameters let us to have a considerable thickness of Cu with a backing temperature of [1297-1318] K, far from 1358K at which Cu melts.

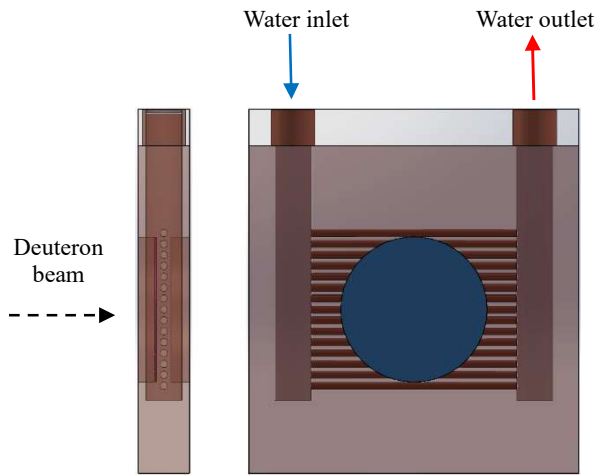

Figure 4. The backing is 7 mm thick, with a width and a height of 45 mm. It is composed of 15 millichannels of 1 mm in diameter with 1.5 mm of separation between them. There is an inlet and outlet for the cooling fluid, with a radius of 2.5 mm. The sample is in a 2.5 mm thick notches, allowing a distance between the $^{176}Yb_2O_3$ sample and the millichannels of 0.5 mm. The both dimensions of ytterbium sample are 10 mm in radius and 0.25 mm in thickness.

Therefore, the final model chosen is a millichannels cooling system, located behind the target to optimize cooling, with a fluid velocity set at 12 m/s. Fig. 4 shows a detailed view of the Cu backing.

The backing is 7 mm thick, with a width and a height of 45 mm. It is composed of 15 millichannels of 1 mm in diameter with 1.5 mm of separation between them. There is an inlet and outlet for the cooling fluid, with a radius of 2.5 mm. The sample is in a 2.5 mm thick notches, allowing a distance between the $^{176}Yb_2O_3$ sample and the millichannels of 0.5 mm. The both dimensions of ytterbium sample are 10 mm in radius and 0.25 mm in thickness.

In addition, to carry out the simulations different parameters were provided to SolidWorks. These parameters were: the water inlet velocity, which has been set at 12 m/s as we specified above, the powers provided by backing and ytterbium to the system, and the contact resistant, which will be studied below. The backing and ytterbium powers has been calculated according to section 2.1 and SRIM [16]. These values are shown in Table 1.

|  | Power (W/m²) |
|---|---|
| Backing | $1.15 \cdot 10^8$ |
| Ytterbium | $1.21 \cdot 10^7$ |

Table 1. Powers provided by both backing with two notches and ytterbium to the system and employed at SolidWorks simulations.

## 4. RESULTS

Once all the variables of our study have been presented, we are going to show the results for different contact resistances. First, we are going to configure the ideal conditions to perform our experiment, zero contact resistance, and we are going to present the results of the production of $^{177}Lu$ obtained with these conditions.

We have performed a study considering a realistic target simulated with SolidWorks. The goal is to fix the most important parameters: maximum current to be sustained at 40 MeV, geometry of the target and cooling system configuration, for a $^{176}Yb_2O_3$ sample of 0.25 mm thickness.

On the other hand, our purpose in this work is to compare the theoretical production of radioisotopes at IFMIF-DONES with deuterons to the current production at nuclear reactors by means of neutron capture reactions. In case of nuclear reactors, the chemical form of the Yb sample is $^{176}Yb_2O_3$ with a maximum possible enrichment of 97% [8], the same target will be taken as objective in this work.

## 4.1 REALISTIC MODEL WITH SOLIDWORKS

Simulations have been carried out for different contact resistances between our backing and the $^{176}$Yb$_2$O$_3$ sample. The initial parameters are the power dissipated by the backing and the sample, the environment conditions, and the cooling water speed of 12 m/s through the cooling system. The inlet water temperature is 278K.

With this model, the 40 MeV deuterons beam is fully stopped in the Cu backing. Furthermore, neither the sample nor the backing melts.

The thermal contact resistance is introduced in the computation procedure when the heat wave propagation reaches the $^{176}$Yb$_2$O$_3$/Cu interface, oxide/metal respectively. With SolidWorks is possible to calculate the critical temperatures for high and low contact resistance values. These resistances are on the order of $10^{-8}$-$10^{-7}$ K·m$^2$/W [20].

Considering that our ytterbium layer has a thickness of 0.25 mm, the most important results are summarized in the following table:

| Contact Resistance (K·m$^2$/W) | Temperature (K) | | |
| --- | --- | --- | --- |
| | $^{176}$Yb$_2$O$_3$ | Cu | Ratio (%) |
| 0 | 1122 | 1122 | 0.00 |
| $1.0 \cdot 10^{-8}$ | 1123 | 1123 | 0.03 |
| $3.0 \cdot 10^{-8}$ | 1122 | 1122 | 0.07 |
| $5.0 \cdot 10^{-8}$ | 1121 | 1120 | 0.11 |
| $8.0 \cdot 10^{-8}$ | 1122. | 1120 | 0.17 |
| $1.0 \cdot 10^{-7}$ | 1123 | 1121 | 0.22 |
| $4.0 \cdot 10^{-7}$ | 1130 | 1121 | 0.84 |
| $5.0 \cdot 10^{-7}$ | 1134 | 1122 | 1.06 |

Table 2. Temperature obtained in the $^{176}$Yb$_2$O$_3$/Cu interface, for different contact resistances between the sample and our backing with two notches.

These simulations have been performed to see what happens when the contact resistance increases by keeping the temperatures of the $^{176}$Yb$_2$O$_3$ sample and the backing below their melting points, 2628 K and 1358 K, respectively. See Table 2.

To study the thermal phenomena which take place inside the $^{176}$Yb$_2$O$_3$/Cu interface it is crucial to estimate the property of contact resistance simultaneously with the thermal conductivity [20]. According to literature and simulated results shown in Table 2, an optimal value for the contact resistance would be in the range of $[10^{-8}$- $4.5 \cdot 10^{-7}]$ K·m$^2$/W, where the $^{176}$Yb$_2$O$_3$/Cu temperature ratio of the interface is lower than 1%.

The temperature distributions for a zero contact resistance are shown in Fig. 5 and Fig. 6.

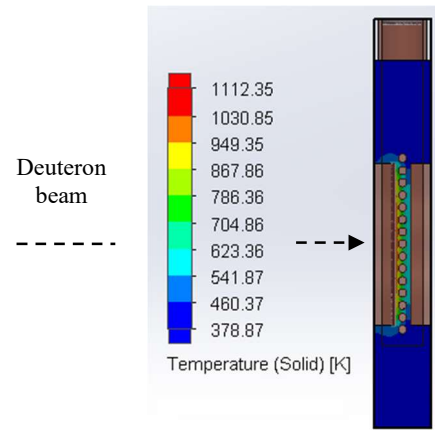

Deuteron beam

Figure 5. Side view for the temperature distribution of the backing and the sample for a cooling water speed of 12 m/s and zero contact resistance.

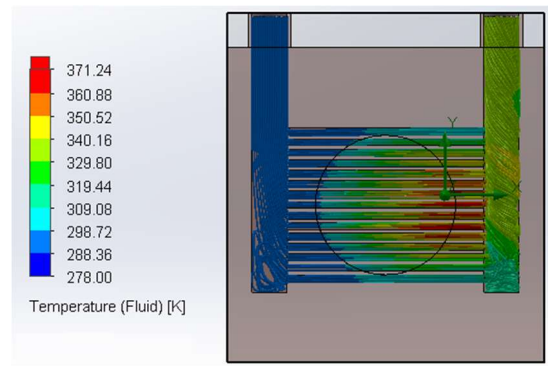

Figure 6. Temperature distribution of the cooling system fluid for a cooling water speed of 12 m/s and zero contact resistance.

## 4.2 ACTIVITY AND MASS PRODUCCED OF $^{177}$LU

The production rates, mass and activity of Yb$_2$O$_3$ have been calculated. Our calculations have kept as boundary condition the temperature of the foil well below their melting points.

For the following results, we have considered the production parameters shown in Table 3. Table 4 shows the results of the production rate of $^{177}$Lu with Yb$_2$O$_3$ sample.

| | |
| --- | --- |
| Deuteron energy | 40 MeV |
| Deuteron current | 1.25 mA |
| Deuteron charge | $1.602 \cdot 10^{-19}$ C |
| Half-live ($^{177}$Lu) | 6.6475 d |
| Mass ($^{177}$Lu) | 176.94 uma |
| Contact Resistance | 0 K·m$^2$/W |

Table 3. Production parameters of $^{177}$Lu.

| | $R_{Lu}$ (mg/s) |
| --- | --- |
| Direct route | $2.26 \cdot 10^{-7}$ |
| Indirect route | $4.82 \cdot 10^{-8}$ |

Table 4. Values obtained for the production rate of $^{177}$Lu in milligram per second, R$_{Lu}$ (mg/s) with Yb$_2$O$_3$ sample.

Then, we have calculated the activity of $^{177}$Lu for 24 hours irradiation.

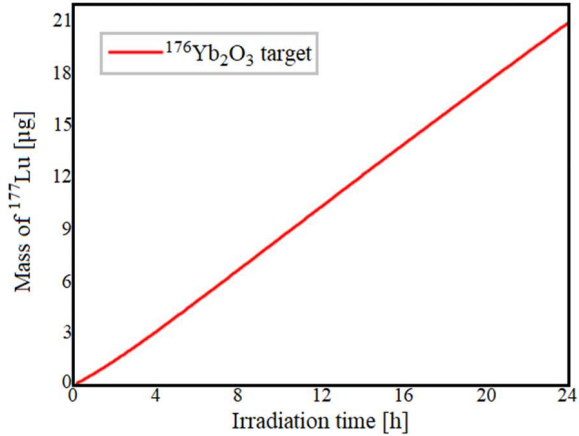

Figure 7. $^{177}$Lu mass produced by $^{176}$Yb$_2$O$_3$ sample. Direct and indirect routes are included in the calculation.

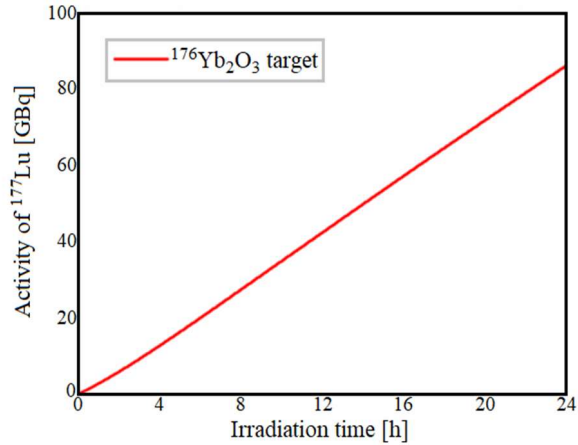

Figure 8. $^{177}$Lu activity produced by $^{176}$Yb$_2$O$_3$ sample. Direct and indirect routes are included in the calculation.

The activity, after 24 hours of irradiation, is 86.11 GBq and the mass of $^{177}$Lu is 20.96 µg. The specific activity of the sample, without purification, is $1.20 \cdot 10^{-4}$ GBq/µg. Note that the production of $^{177m}$Lu, according to Hermanne studies [17], does not exceed 0.0045%.

Finally, table 5 shows the comparative between the activity generated with neutrons at nuclear reactors, and the activity generated with deuterons at IFMIF-DONES. It is observed that more activity is generated in much less time. In the bibliography it is found that one can reach, with neutrons at nuclear reactors, an specific activity of 738 GBq/mg by direct route, $^{176}$Lu(n,γ)$^{177}$Lu+$^{177m}$Lu, and almost 1.11 GBq/mg by indirect route, $^{176}$Yb(n,γ)$^{177}$Yb→$^{177}$Lu, before the purification of the sample.

In these theoretical results obtained in IFMIF-DONES, a purification of the sample has not been considered as in the data presented for the experiments in the reactors. It should be noted that there are several purification methods that must be performed after the experiment as shown in [8].

| Reaction | | Irradiation time (d) | Activity (GBq) | Specific Activity (GBq/mg) |
|---|---|---|---|---|
| $^{176}$Lu(n,γ)$^{177}$Lu+$^{177m}$Lu | | 8 | -- | 738 |
| $^{176}$Yb(n,γ)$^{177}$Yb→$^{177}$Lu | | 7 | -- | 1.11 |
| $^{176}$Yb | (d,n/p)$^{177(m+g)}$Lu | 1 | 86.11 | 0.12 |
| | | 8 | 510.20 | 0.71 |

Table 5. Comparative between the activity generated with neutrons reactions and the activity generated with deuterons on IFMIF-DONES.

## 5. DISCUSSION AND CONCLUSIONS

We have studied the production of $^{177}$Lu, used as a theranostics isotope, through the two specific production routes, with deuterons of 40 MeV at IFMIF-DONES.

Previous study shows that the intensity of the deuteron beam should be reduced from 125 mA to 1.25 mA. From here, a realistic model has been carried out in SolidWorks, taking zero contact resistance and inlet fluid velocity of 12 m/s, in order to find the optimal dimensions of the geometry that provide temperatures for both the backing of Cu and the $^{176}$Yb$_2$O$_3$ sample lower than its melting point, maximizing the activity produced.

Additionally, the influence of the variability of the contact thermal resistance parameter, which is an extrinsic characteristic of materials, has been studied. Contact thermal resistance decreases because of the increase contact pressure at the interface or due to the decrease roughness at the same interface, ensuring higher heat transfer.

Then, the activity and the mass generated of $^{177}$Lu during 24 hours of irradiation have been calculated. Lastly, the specific activity obtained after 24 h of IFMIF-DONES irradiation is 0.116 GBq/mg for the $^{176}$Yb$_2$O$_3$ sample. We can compare these values with the specific activity produced in nuclear reactors with a $^{176}$Yb$_2$O$_3$ sample after irradiation for 72 h, the result of which is 2.89 TBq/mg [8]. It should be noted that Hermanne's studies [17], although they do not give exact data, indicate that the specific activity by reaction with deuterons is quite high. Although we cannot quantify the specific activity of the sample, the perspective, according to the previous approach, is promising for the specific activity that can be reached in IFMIF-DONES.

Furthermore, the production at IFMIF-DONES would have a considerable impact in a regional health system as in Granada (Spain). Theragnosis of neuroendocrine tumours has recently experienced a breakthrough thanks to lutetium radiopharmaceuticals [9]. A complete treatment for a patient requires 4 doses. Each dose has a price of 14 k€ and an activity of 7.4 GBq. Therefore, in case of 24 h irradiation at IFMIF-DONES, the produced $^{177}$Lu could bring to save about 112 k€. New lutetium radiopharmaceuticals are emerging, as $^{177}$Lu-PSMA 617 [10] for prostate cancer therapy, breast cancer [11] and Lu-labelled-antibodies [12], so the growing interest in $^{177}$Lu production must be considered. It should be stressed that we are only providing preliminary calculations with the aim to motivate more realistic studies.

**Acknowledgments.** This work was carried out within the framework of the EUROfusion Consortium and has received funding from the Euratom research and training programme 2014-2018 and 2019-2020 under grant agreement No 633053. The views and opinions expressed herein do not necessarily reflect those of the European Commission. This work was supported by DONES-PreP project (870186), the Spanish projects A-FQM-371-UGR18 (Programa Operativo FEDER Andalucía 2014-2020), P20_00665 (Proyectos I+D+i Junta de Andalucia 2020), PID2020-117969RB-I00 (Proyectos del Plan Nacional 2020) and the sponsors of the University of Granada Chair Neutrons for Medicine: Fundación ACS, Capitán Antonio and La Kuadrilla.

[1] F. Mota et al., *Sensitivity of IFMIF-DONES irradiation characteristics to different design parameters*, Nucl. Fusion 55, 12 (2015).

[2] http://www.roadmap2018.esfri.eu/projects-and-landmarks/browse-the-catalogue/ifmif-dones/

[3] DONES-PreP. Grant agreement ID: 870186. https://cordis.europa.eu/project/id/870186.

[4] W. Królas et al., *The IFMIF-DONES fusion oriented neutron source : evolution of the design,* Nucl. Fusion (2021). https://doi.org/10.1088/1741-4326/ac318f

[5] J. Knaster et al., *IFMIF, the European-Japanese efforts under the Broader Approach agreement towards a Li(d,xn) neutron source: Current status and future options*, Nucl. Materials and Energy 9, 46-54 (2016).

[6] A. Maj et al., *White book on the complementary scientific programme at IFMIF-DONES*. http://www.ifj.edu.pl/publ/reports/2016/

[7] J. Praena et al., ENS-7.2.3.1-T13-06-N1 Feasibility study of the use of DONES for Radioisotope Production, Electronics Irradiation and Neutron Scattering, 21 January (2020).

[8] A. Dash et al., *Production of $^{177}$Lu for Targeted Radionuclide Therapy: Available Options*, Nucl. Med. Mol. Imaging 49, 85-107 (2015).

[9] A.T. Kendi et al., *Therapy With 177Lu-DOTATATE: Clinical Implementation and Impact on Care of Patients With Neuroendocrine Tumors*, Nucl. Med. Mol. Imaging 213, 2 (2019).

[10] C. Kratochwil et al., *EANM procedure guidelines for radionuclide therapy with 177Lu-labelled PSMA-ligands (177Lu-PSMA-RLT)*, Eur. J. Nucl. Med. Mol. Imaging 46 (12), 2536-2544 (2019).

[11] A.C. Camargo Miranda et al., *Radioimmunotheranostic Pair Based on the Anti-HER2 Monoclonal Antibody:Influence of Chelating Agents and Radionuclides on Biological Propoerties,* Pharmaceutics (2021), 13, 971. https://doi.org/10.3390/pharmaceutics13070971

[12] A. Ku et al., *Dose predictions for [$^{177}$Lu]Lu-DOTA-panitumumab F(ab')$_2$ in NRG mice with HNSCC patient-derived tumour xenografts based on [$^{64}$Cu]Cu-DOTA-panitumumab F(ab')$_2$– implications for a PET theranostic strategy,* EJNMMI Radiopharmacy and Chemistry (2021) 6:25. https://doi.org/10.1186/s41181-021-00140-1

[13] M.R.A. Pillai et al., *Production logistics of $^{177}$Lu for radionuclide therapy*, Appl. Rad. Iso. 59 (2-3), 109–118 (2003).

[14] J. Praena et al., *Radioisotope production at the IFMIF-DONES facility*, EPJ Web of conferences 239, 23001 (2020). https://doi.org/10.1051/epjconf/202023923001

[15] H.H. Andersen and J.F. Ziegler, *Stopping power and ranges in all elements*, Pergamon Press, New York, 1985.

[16] http://www.srim.org/

[17] A. Hermanne et al., Nucl. Inst. and Met. in Phys. Res. B. 247 (2006) 223-231.

[18] S. Manenti et al., Appl. Rad. Iso. 69 (2011) 37–45.

[19] F. Arias de Saavedra et al., *Routes for the production of isotopes for PET with high intensity deuteron accelerators*, Nucl. Inst. and Met. in Phys. Res A 887 (2018) 50-53.

[20] S.Orain et al., *Use of genetic algorithms for the simultaneous estimation of thin films thermal conductivity and contact resistances*, International Journal of Heat and Mass Transfer 44 (2001) 3973-3984